\documentclass[a4paper,12pt]{article}
\usepackage{latexsym}
\usepackage{amsmath,amssymb,bm,ascmac,color,calc,subfigure}
\usepackage[mathscr]{eucal}
\usepackage{subfigure}
\usepackage{graphicx}
\usepackage{cite}
\usepackage{lineno}


\def\sla#1{\rlap{\kern .15em /}#1}

\begin{document}
\begin{titlepage}

\begin{flushright}
\begin{tabular}{l}
KEK-TH-1715
\end{tabular}
\end{flushright}

\vspace*{2cm}

\begin{center}
{\large\bf 
CP violation in {\boldmath $B\to D\tau \nu_{\tau}$} using multi-pion tau decays}\\[15mm]
Kaoru Hagiwara\footnote{E-mail: kaoru.hagiwara@kek.jp}$^{,a}$, 
Mihoko M. Nojiri\footnote{E-mail: nojiri@post.kek.jp}$^{,a,b}$, and 
Yasuhito Sakaki\footnote{E-mail: sakakiy@post.kek.jp}$^{,a}$\\ \bigskip
{\em 
$^{a}$KEK Theory Center and Sokendai, \\
Tsukuba, Ibaraki 305-0801, Japan\\ \bigskip
$^{b}$Kavli IPMU (WPI), The University of Tokyo,\\
Kashiwa, Chiba 277-8583, Japan.}
\\[20mm] 
\end{center}

\begin{abstract}
Present experimental datas have shown a 3.8$\sigma$ level discrepancy with the standard model in $\overline{B}\to D^{(*)}\tau\bar{\nu}_{\tau}$. 
Some new physics models have been considered to explain this discrepancy possibly with new source of the CP violation. 
In this paper, we construct CP violating observables by using multi-pion decays in $B\to D\tau\nu_{\tau}$, 
and estimate sensitivity of these observables to generic CP violating operators. 
We also discuss possibilities of CP violation in leptoquark models and in 2HDM of type-III. 
\end{abstract}
\end{titlepage}

\renewcommand\thefootnote{\roman{footnote}}
\section{Introduction}
The standard model (SM) gives an accurate description of elementary particle phenomena, however, experimenatal uncertainties about the flavor structure of the third generation are still larger than the first and second generations. 
In the standard model, the charged currents are described by the $SU(2)_L$ gauge coupling of the left-handed doublets of quarks and leptons, and the Cabibbo-Kobayashi-Maskawa flavor-mixing matrix. 
Some models beyond the standard model predict different structures. 
A typical example are two Higgs doublet models (2HDMs), 
which predict charged Higgs contributions, 
and its couplings are proportional to the fermion masses. 

For testing the universality of charged currents among the lepton generations, ratios of the branching fractions are introduced as observables,
\begin{align}
R(D^{(*)})
=
\frac
{{\rm Br}(\overline{B}\to D^{(*)}\tau\bar{\nu})}
{{\rm Br}(\overline{B}\to D^{(*)}\ell\bar{\nu})},
\end{align}
where $\ell$ denotes $e$ or $\mu$. 
The standard model predictions are given in the Refs. 
\cite{Nierste:2008qe,Kamenik:2008tj,Tanaka:2010se,Fajfer:2012vx,Sakaki:2012ft,Bailey:2012jg,Becirevic:2012jf}.  
These values in Refs. \cite{Tanaka:2012nw,Sakaki:2013bfa} are, 
\begin{align}
R(D)_{\rm SM}=0.305\pm 0.012,~~R(D^*)_{\rm SM}=0.252\pm 0.004. \label{SMpred}
\end{align}
Predictions of the minimal supersymmetric standard model (MSSM) are found 
in Refs. \cite{Hou:1992sy,Tanaka:1994ay,Kamenik:2008tj,Nierste:2008qe,Tanaka:2010se}, 
which can significantly affect the semi-tauonic B decays, 
through its Higgs sector of the type-II 2HDM. 

The present experimental datas are given by the BaBar Collaboration \cite{Lees:2012xj,Lees:2013uzd}:
\begin{align}
R(D)_{\rm BaBar}=0.440\pm 0.072,~~R(D^*)_{\rm BaBar}=0.332\pm 0.030, \label{BaBar_exp}
\end{align}
with the error correlation of $\rho = -0.27$. 
These results are inconsistent with the SM  of Eq. (\ref{SMpred}) at 3.4$\sigma$, or 99.93\% CL for the two data points. 
The type II 2HDM does not improve the fit, being inconsistent with the data at 99.8\% CL for the optimal value of $m_{H^{\pm}}/\tan\beta$. 
The Belle Collaboration also reported  measurements \cite{Matyja:2007kt,Adachi:2009qg,Bozek:2010xy}, 
and the newest results which are estimated in Ref. \cite{Lees:2013uzd} are, 
\begin{align}
R(D)_{\rm Belle}=0.34 \pm 0.12 ,~~~R(D^*)_{\rm Belle}=0.43 \pm 0.08, \label{Belle_exp}
\end{align}
where the error correlation value is not given. If we assume the same negative correlation of $\rho=-0.27$, 
the BaBar (\ref{BaBar_exp}) and Belle (\ref{Belle_exp}) data can be combined to give
\begin{align}
R(D)_{\rm exp}=0.42 \pm 0.06,~~~R(D^*)_{\rm exp}=0.34 \pm 0.03, \label{comb_exp}
\end{align}
with the error correlation of $\rho = -0.26$.\footnote[1]{If the error correlation is set to zero in the Belle data (\ref{Belle_exp}), the combined results give $R(D)_{\rm exp}=0.41 \pm 0.06$ and $R(D^*)_{\rm exp}=0.34 \pm 0.03$ with the correlation $\rho=-0.22$, which are inconsistent with the SM at 3.7$\sigma$ (99.976\% CL). }
The combined data is now inconsestent with the SM prediction (\ref{SMpred}) at 3.8$\sigma$ (99.985\% CL). 

Some new physics scenarios consistent with the datas have been considered in Refs. 
\cite{Tanaka:2012nw,Fajfer:2012vx,Becirevic:2012jf,Sakaki:2012ft,Datta:2012qk,Bailey:2012jg,Fajfer:2012jt,Celis:2012dk,Biancofiore:2013ki,Ko:2012sv,Celis:2013jha}. 
In the articles \cite{Sakaki:2013bfa,Dorsner:2013tla}, 
all leptoquark models compatible with the SM gauge group are studied.  
These models are found to be able to explain the experimental results. 
The 2HDM of type III can also account for the discrepancy \cite{Crivellin:2012ye}. 
In Ref. \cite{Tanaka:2012nw}, model independent analysis has been performed, 
and the Wilson coefficients of generic dimension-6 operators which are favored by the experimental data are identified. 
The allowed regions of the coefficients reside on the complex plane, 
and the imaginary parts induce CP violation. 
In Ref. \cite{Duraisamy:2013pia}, some observables which are sensitive to CP violation are constructed by using the $D^*$ polarization, which is difficult to measure.  

In this paper, we discuss CP violating observables involving the polarization of vector resonances $(\rho,a_1,{\rm etc.})$
produced in the decay of the tau lepton in $B\to D^{(*)}\tau\nu$. 
In Sec. \ref{Formalism}, we construct observables which have sensitivity to CP violation in a general situation, 
and only consider $B\to D\tau\nu$ for simplicity.  
In Sec. \ref{Numerical}, we examine the sensitivities of these observables to two independent imaginary parts of new physics Wilson coefficients. 
Furthermore, we examine CP violation term in three leptoquark models, and 2HDM of type-III. 
Finaly, we summarize our findings in Sec. \ref{Conclusion}.

\section{Formalism}
\label{Formalism}
\subsection{Effective Hamiltonian and amplitudes}
Assuming the tau neutrinos to be left-handed, we introduce a general effective Hamiltonian that contains all possible four-fermion operators of the lowest dimension for the $b\to c\tau^-\bar{\nu}_{\tau}~(\bar{b}\to \bar{c}\tau^+\nu_{\tau}) $ transition \cite{Tanaka:2012nw},
\begin{align}
\mathcal{H}_{\rm eff}=\frac{4G_F}{\sqrt{2}}V_{cb}
[
(1+C_{V_1})\mathcal{O}_{V_1}
+C_{V_2}\mathcal{O}_{V_2}
+C_{S_1}\mathcal{O}_{S_1}
+C_{S_2}\mathcal{O}_{S_2}
+C_{T}\mathcal{O}_{T}
]+{\rm H.c.},
\end{align}
where the four-Fermi operators are defined as
\begin{align}
\mathcal{O}_{V_1} &=(\bar{c}_L\gamma^{\mu}b_L)(\bar{\tau}_L\gamma_{\mu}\nu_{\tau L}),\\
\mathcal{O}_{V_2}&=(\bar{c}_R\gamma^{\mu}b_R)(\bar{\tau}_L\gamma_{\mu}\nu_{\tau L}),\\
\mathcal{O}_{S_1}&=(\bar{c}_L b_R)(\bar{\tau}_R \nu_{\tau L}),\\
\mathcal{O}_{S_2}&=(\bar{c}_R b_L)(\bar{\tau}_R \nu_{\tau L}),\\
\mathcal{O}_{T}&=(\bar{c}_R\sigma^{\mu\nu} b_L)(\bar{\tau}_R \sigma_{\mu\nu} \nu_{\tau L}).
\end{align}
We consider the decay process $\overline{B}\to D \tau \overline{\nu}_{\tau}$ followed by the tau lepton decay
into two or three pions via a vector resonance, 
\begin{subequations}
\label{katei}
\begin{align}
&\overline{B}(p_{B})\longrightarrow D(p_D) \tau^-(p_{\tau}) \overline{\nu_{\tau}}(p_{\nu_1})\\[4pt]
&\hspace{4.1cm}V^-(Q_{2,3})\nu_{\tau}(p_{\nu_2})\\
&\hspace{5.2cm}\pi^-(p_{1})\pi^0(p_{2}) \label{pro_2pi}\\
&\hspace{5.2cm}\pi^+(p_{1})\pi^-(p_{2})\pi^-(p_{3}) \label{pro_3pi_1}\\
&\hspace{5.2cm}\pi^-(p_{1})\pi^0(p_{2})\pi^0(p_{3}) \label{pro_3pi_2}
\begin{picture}(0,0)(127,-80)
\put(-24,-6){\line(0,-1){13}}
\put(-25,-22){$\longrightarrow$}
\put(  5,-28){\line(0,-1){11}}
\put(  4,-42){$\longrightarrow$}
\end{picture}\\[-20pt]
\nonumber
\end{align}
\end{subequations}
where $V$ denotes vector resonances. The vector resonances are $\rho$ and $\rho '$ meson for the 2$\pi$ decay (\ref{pro_2pi}) and $a_1$ meson for 3$\pi$ decay (\ref{pro_3pi_1},\ref{pro_3pi_2}), 
whose decay branching fractions are about 26\% and 18\%, respectively. 
In the case of 3$\pi$ decay, we assign the momentum $p_1$ to a pion which have a different electric charge from the  others two pions. 
In this paper, $\overline{B}\to D$ denotes $\overline{B^0}\to D^+$ or $B^- \to D^0$ transitions. 

The differential decay rate for the processes  
$\overline{B}
\to D \tau^- \overline{\nu_{\tau}} 
\to D \nu_{\tau}\overline{\nu_{\tau}} 
+ n\pi 
$ 
(\ref{katei}) are written as
\begin{align}
d\Gamma_n=\sum_{\lambda \lambda'}
d\Gamma_V^{\lambda \lambda'}
|F_n(Q_n^2)|^2
\frac{dQ_n^2}{2\pi}
d\Gamma_n^{\lambda \lambda'},
\label{differential_decay_rate_Bbar}
\end{align}
where $n$ is a pion number in the process, $Q_n$ denotes the 4-momentum of vector resonance, $Q_n=\sum_{i=1}^n p_i$. 
The density matrix distribution of the vector resonance production can be expressed as\footnote[2]{
As explained in the following section, we perform all the numerical analysis by using amplitudes in TAUOLA \cite{Jadach:1990mz,Jezabek:1991qp}, which takes account of the scalar $(\pi')$ contribution to the 3$\pi$ mode. 
Although the scalar contributions can be included in our formalism by introducing the scalar polarization $(\lambda=4)$ of the ``vector" resonance, 
we keep the 3$\times$3 form of the density matrix (with $\lambda,\lambda'=\pm 1,0$) for the sake of brevity. 
None of our final results are sensitive to the $\pi'$ contribution to the 3$\pi$ channel. }
\begin{align}
d\Gamma_V^{\lambda \lambda'}
&=
\frac{1}{2 m_B}d\Phi_4 (\overline{B}\to D V \nu\bar{\nu }) \mathcal{M}_V^{\lambda }\mathcal{M}_V^{\lambda'*}, 
\end{align}
where $d\Phi_4 (\overline{B}\to D V \nu\bar{\nu })$ denotes the invariant 4-body phase space for the vector boson of the momentum squared $\sqrt{Q_n^2}$, 
and the decay matrix elements
\begin{align}
\mathcal{M}_V^{\lambda}
&=
(\sqrt{2}G_F)^2 V_{cb}\cos\theta_C
	\bar{u}(p_{\nu_2})P_{+}\gamma^{\alpha}
	\frac{\sla{p}_{\tau}+m_{\tau}}{p_{\tau}^2-m_{\tau}^2+i m_{\tau}\Gamma_{\tau}}
	\Gamma_{\rm NP}P_{-}v(p_{\nu_1})
	\epsilon_{\alpha}^*
	(Q_n,\lambda), 
	\label{decay_matrix_elements}
\end{align}
are given for the vector boson of helicity $\lambda$ in the rest frame of the $\tau\overline{\nu}$ system, where
\begin{align}
\Gamma_{\rm NP}
&=
C_V H_V^{\mu}\gamma_{\mu}+C_S H_S+C_T H_T^{\mu\nu}\sigma_{\mu\nu} \label{GammaNP}\\
&=
e^{i {\rm arg}(C_V)}(|C_V| H_V^{\mu}\gamma_{\mu}+\widetilde{C}_S H_S+\widetilde{C}_T H_T^{\mu\nu}\sigma_{\mu\nu}) 
\end{align}
give the $\overline{B}\to D \tau \bar{\nu}_{\tau}$ matrix elements of the general $b\to c \tau \bar{\nu}_{\tau}$ contact interactions with new physics contributions. 
We parametrize the three Wilson coefficients as
\begin{align}
C_V &=1+C_{V_1}+C_{V_2}
~,~~C_{S}=C_{S_1}+C_{S_2}
~,~~\widetilde{C}_{S,T}=e^{-i {\rm arg}(C_V)} C_{S,T}, \label{CXpara}
\end{align}
where $C_V =1$ stand for the SM, and by factoring out the overall phase of the generalized coefficients $C_V$. 
In Eq. (\ref{CXpara}), $P_{\pm}=(1 \pm \gamma_5)/2$, $\theta_C$ denotes the Cabbibo angle, 
and $\epsilon^{\alpha}(Q,\lambda)$ is the polarization vector of vector resonances. 
The hadronic amplitudes $H_{V,S,T}$ are defined as 
\begin{align}
H_V^{\mu}
&=
\langle D(p_D) |\bar{c}\gamma^{\mu}b| \overline{B}(p_B) \rangle , \label{HV}\\
H_S
&=
\langle D(p_D) |\bar{c}b| \overline{B}(p_B) \rangle, \\
H_T^{\mu\nu}
&=
\langle D(p_D) |\bar{c}\sigma^{\mu\nu}(1-\gamma_5)b| \overline{B}(p_B) \rangle .\label{HT}
\end{align}
The pseudo-vector and the pseudo-scalar hadronic amplitudes are zero,
\begin{align}
\langle D|\bar{c}\gamma^{\mu}\gamma_5 b| \overline{B} \rangle
=\langle D|\bar{c}\gamma_5 b| \overline{B} \rangle=0,
\end{align}
due to the parity conservation in the strong interactions. 
Three hadronic amplitudes in Eqs. (\ref{HV}-\ref{HT}) have been parameterized by using the heavy quark effective theory \cite{Caprini:1997mu}, 
and measured by the BaBar and Belle experiments \cite{Aubert:2007rs,Aubert:2008yv,Abe:2001yf,Dungel:2010uk}. 
We adopt the parametrization given in the Ref. \cite{Sakaki:2013bfa} in our numerical analysis. 

In Eq. (\ref{differential_decay_rate_Bbar}), the density matrix of the $V\to n \pi$ decay distribution is defined as 
\begin{align}
d\Gamma_n^{\lambda \lambda'}
&=
d\Phi_n \mathcal{M}_n^{\lambda}\mathcal{M}_n^{\lambda'*}, 
\end{align}
where the $V\to n \pi$ decay amplitudes $\mathcal{M}_n^{\lambda}$ are given in Refs. \cite{Jadach:1990mz,Jezabek:1991qp,Hagiwara:2012vz}. 
For the 2$\pi$ decay $V^{-}\to \pi^{-}(p_1)\pi^{0}(p_2)$, 
\begin{align}
\mathcal{M}_2^{\lambda}
&=
\sqrt{2} \epsilon^{\alpha}(Q_2,\lambda)(p_1-p_2)_{\alpha} \label{M2}, 
\end{align}
and for the 3$\pi$ decay 
$V^{-}\to \pi^{+}(p_1)\pi^{-}(p_2)\pi^{-}(p_3) \text{ or } \pi^{-}(p_1)\pi^{0}(p_2)\pi^{0}(p_3)$, 
\begin{align}
\mathcal{M}_3^{\lambda}
&=
\frac{4}{3f_{\pi}} \epsilon^{\alpha}(Q_3,\lambda)[
  F_2(P_{12}^2)(p_1-p_2)_{\alpha}
+F_2(P_{13}^2)(p_1-p_3)_{\alpha}
], \label{M3}
\end{align}
with 
\begin{align}
P_{ij}=p_i+p_j  \label{Pij}. 
\end{align} 
The form factors $F_n(Q_n^2)$ are parametrized in Refs. \cite{Jadach:1990mz,Jezabek:1991qp} as 
\begin{align}
F_{2}(Q^2)&=[B_{\rho}(Q^2)+\alpha B_{\rho'}(Q^2)]/(1+\alpha), \label{F2def}\\
F_{3}(Q^2)&= B_{a_1}(Q^2)\label{F3def}, 
\end{align}
by using a modified Breit-Wigner propagator $B_{V}(Q^2)$ 
\begin{align}
B_{V}(Q^2)&=\frac{m_V^2}{m_V^2-Q^2-i\sqrt{Q^2}\Gamma_V(Q^2)}. \label{breit}
\end{align}
We consider $\rho$ and $\rho'$ resonances in the case of $V\to2\pi$ decay, with  $\alpha = -0.145$ \cite{Jadach:1990mz,Jezabek:1991qp} in Eq. (\ref{F2def}). 
The runnnig width is defined as 
\begin{align}
\sqrt{Q^2} \Gamma_V(Q^2)
&=m_V \Gamma_V
\frac{Q^2 g_V(Q^2)}{m_V^2 g_V(m_V^2)}, 
\end{align}
where the line shape factors are
\begin{align}
g_{\rho, \rho'}(Q^2)
&=
\bar{\beta}\left( \frac{m_{\pi^-}^2}{Q^2} , \frac{m_{\pi^0}^2}{Q^2} \right),\\
g_{a_1}(Q^2)
&= \begin{cases}
  \frac{4.1}{Q^2}(Q^2-9m_{\pi}^2)^3[1-3.3(Q^2-9m_{\pi}^2)+5.8(Q^2-9m_{\pi}^2)^2] 
  	& \text{if}~Q^2<(m_{\rho}+m_{\pi})^2 \\
  1.623+\frac{10.38}{Q^2}-\frac{9.32}{Q^4}+\frac{0.65}{Q^6} 
  	& \text{if}~Q^2>(m_{\rho}+m_{\pi})^2
  \end{cases}
\end{align}
with
\begin{align}
\bar{\beta}(a,b)=(1+a^2+b^2-2a-2b-2ab)^{1/2}. 
\end{align}
%

\subsection{CP asymmetries}
We consider $B\to \overline{D}$ processes for comparing with $\overline{B}\to D$ processes in Eq. (\ref{katei}):
\begin{subequations}
\label{CPkatei}
\begin{align}
&B(p_{B})\longrightarrow \overline{D}(p_D) \tau^+(p_{\tau}) \nu_{\tau}(p_{\nu_1})\\[4pt]
&\hspace{4.1cm}V^+(Q_{2,3})\overline{\nu_{\tau}}(p_{\nu_2})\\
&\hspace{5.2cm}\pi^+(p_{1})\pi^0(p_{2})\\
&\hspace{5.2cm}\pi^-(p_{1})\pi^+(p_{2})\pi^+(p_{3})\\
&\hspace{5.2cm}\pi^+(p_{1})\pi^0(p_{2})\pi^0(p_{3})
\begin{picture}(0,0)(127,-80)
\put(-24,-6){\line(0,-1){13}}
\put(-25,-22){$\longrightarrow$}
\put(  5,-28){\line(0,-1){11}}
\put(  4,-42){$\longrightarrow$}
\end{picture}\\[-20pt]
\nonumber
\end{align}
\end{subequations}
In this paper, $B\to \overline{D}$ denotes $B^0\to D^-$ or $B^+ \to \overline{D^0}$ transitions. 
The differential decay rate for these processes are written as 
\begin{align}
d\overline{\Gamma}_n=\sum_{\lambda \lambda'}
d\overline{\Gamma}_V^{\lambda \lambda'}
|F_n(Q_n^2)|^2
\frac{dQ_n^2}{2\pi}
d\Gamma_n^{\lambda \lambda'}, 
\end{align}
where $d\overline{\Gamma}_V^{\lambda \lambda'}$ are given as
\begin{align}
d\overline{\Gamma}_V^{\lambda \lambda'}
&=
\frac{1}{2 m_B}
	d\Phi_4 (B\to \overline{D} V \bar{\nu}\nu)
	\overline{\mathcal{M}}_V^{\lambda}\overline{\mathcal{M}}_V^{\lambda'*}, \\
\overline{\mathcal{M}}_V^{\lambda}
&=
(\sqrt{2}G_F)^2 V_{cb}^* \cos\theta_C
	\bar{u}(p_{\nu_1})P_{+}
	\overline{\Gamma}_{\rm NP}
	\frac{-\sla{p}_{\tau}+m_{\tau}}{p_{\tau}^2-m_{\tau}^2+i m_{\tau}\Gamma_{\tau}}
	\gamma^{\alpha}
	P_{-}v(p_{\nu_2})
	\epsilon_\alpha^{*}
	(Q_n,\lambda), \\
\overline{\Gamma}_{\rm NP}
&=
C_{V}^* H_V^{\mu}\gamma_{\mu} -C_{S}^* H_S-C_T^* H_T^{\mu\nu*}\sigma_{\mu\nu}\\
&=
e^{-i {\rm arg}(C_V)}(
	|C_V^*| H_V^{\mu}\gamma_{\mu}
	-\widetilde{C}_S^* H_S
	-\widetilde{C}_T^* H_T^{\mu\nu*}\sigma_{\mu\nu}) 
\end{align}
From these equations, we obtain the following basic formula, 
\begin{align}
d\Gamma_n
=(d\overline{\Gamma}_n^{\rm P}
~{\rm with}~
(C_V^{*},C_S^{*},C_T^{*})\to (C_V,C_S,C_T)) \label{impCP}, 
\end{align}
where $d\overline{\Gamma}_n^{\rm P}$ denote $d\overline{\Gamma}_n$ with the parity transformation, that is, with reversing the directions of all 3-momenta. 
In short, the distributions are CP invariant if all the Wilson coefficients are real. 

Let us write the total amplitude, 
$\mathcal{M} = \sum_{\lambda} \mathcal{M}_V^{\lambda} \mathcal{M}_n^{\lambda}$, 
the product of the production (\ref{decay_matrix_elements}) and decay 
(\ref{M2},\ref{M3}) amplitudes formally as
\begin{align}
\mathcal{M}=\mathcal{M}_{\rm SM}+C_{\rm NP}\mathcal{M}_{\rm NP}. 
\end{align}
Here $\mathcal{M}_{\rm SM}$ is the amplitude from the SM, $C_{\rm NP}\mathcal{M}_{\rm NP}$ is the amplitude from the new physics, where $C_{\rm NP}$ denote the Wilson coefficients of the new physics $C_V, C_S$ and $C_T$. 
From Eq. (\ref{impCP}), we find 
\begin{align}
d\Gamma_n-d\overline{\Gamma}_n^{\rm P}
\propto
{\rm Im}(C_{\rm NP}){\rm Im}(\mathcal{M}_{\rm SM}^*\mathcal{M}_{\rm NP}). \label{CPAsy_relation}
\end{align}
Therefore, we can measure the imaginary parts of the Wilson coefficients, 
which are the source of the CP violation by using this asymmetry. 

\subsection{CP violating observable in two-pion decay}
\label{CP_violating_observable_in_two-pion_decay}
\begin{figure}[t]
\begin{center}
\includegraphics[width=9.0cm, bb=0 0 435 292]{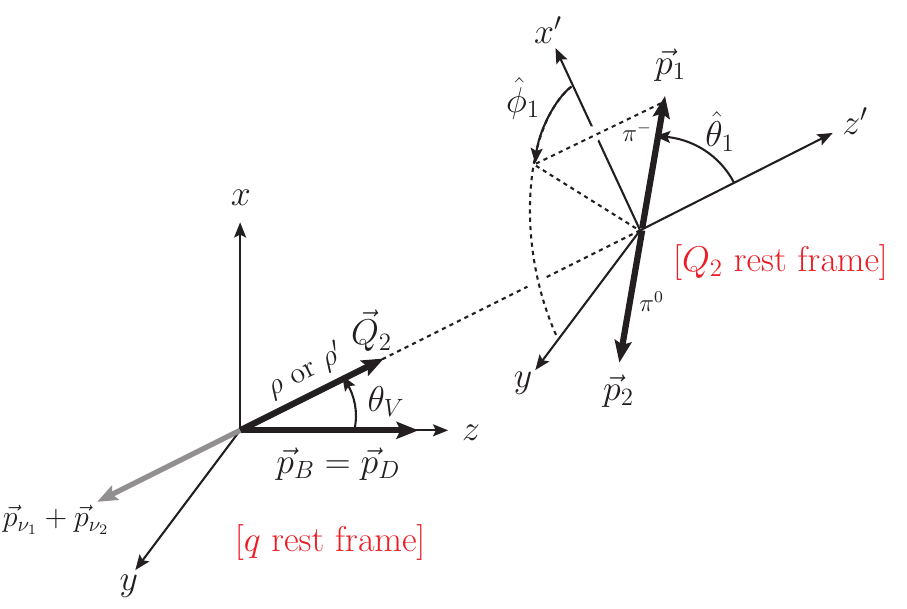}
\caption{{\footnotesize The kinematics in the case of $V\to 2\pi$.}}
\label{n2kin}
\end{center}
\end{figure}
We parametrize the momenta $p_B$, $p_D$, and $Q_2$ in the $q$ rest frame ($ \vec{p}_B-\vec{p}_D=\vec{0}$) as
\begin{align}
p_B&=\frac{\sqrt{q^2}}{2}(\frac{m_B^2-m_D^2}{q^2}+1 ,0,0,\beta), \label{pBdef2pi} \\
p_D&=\frac{\sqrt{q^2}}{2}(\frac{m_B^2-m_D^2}{q^2}-1  ,0,0,\beta), \\
Q_2&=(E_V,|\vec{p}_V|\sin\theta_V,0,|\vec{p}_V|\cos\theta_V), \label{Q2def2pi} 
\end{align}
where $q=p_B-p_D$, $Q_2 = p_{1} + p_{2}$, $E_V$ is the sum of two pion energies, and 
$\beta    =\bar{\beta}\left( \frac{m_B^2}{q^2},\frac{m_D^2}{q^2} \right)$. 
In this frame, we define $z$-axis along the direction of $\vec{p}_D$, 
and $y$-axis along the direction of $\vec{p}_D \times \vec{Q}_2$. 
The opening angle between $\vec{p}_D$ and $\vec{Q}_2$ is denoted by $\theta_V$. 
The two pion momenta $p_1$ and $p_2$ in the $Q_2$ rest frame $( \vec{p}_1+\vec{p}_2=\vec{0})$ can then be parametrized as 
\begin{align}
p_1&=\frac{\sqrt{Q_2^2}}{2}
   (1,+\beta_1 \sin\hat{\theta}_1\cos\hat{\phi}_1,+\beta_1 \sin\hat{\theta}_1\sin\hat{\phi}_1,+\beta_1 \cos\hat{\theta}_1), 
   \label{p1def}\\
p_2&=\frac{\sqrt{Q_2^2}}{2}
   (1,-\beta_1 \sin\hat{\theta}_1\cos\hat{\phi}_1,-\beta_1 \sin\hat{\theta}_1\sin\hat{\phi}_1,-\beta_1 \cos\hat{\theta}_1), 
\end{align}
where 
$\beta_1=\sqrt{ 1-\frac{4m_{\pi}^2}{Q_2^2} }$, 
$z'$-axis is chosen along the direction of $\vec{Q}_2$ in the $q$ rest frame, 
and $y'$-axis remains that of the $y$-axis. 
The polar and azimuthal angle of $\vec{p}_1$, $\hat{\theta}_1$ and $\hat{\phi}_1$, 
are parameterized as shown in Fig. \ref{n2kin}. 
We define the polarization vectors of the vector resonance in this $Q_2$ rest frame as
\begin{align}
\epsilon(\pm)&= \frac{1}{\sqrt{2}}(0,\mp1,-i,0), \\
\epsilon(0)&=(0,0,0,1),
\end{align}
and the density matrix distribution of the vector resonance decay in the case of $V\to 2\pi$ is
\begin{align}
\frac{d\Gamma_2^{\lambda\lambda'}}{d\cos\hat{\theta}_1 d\hat{\phi}_1}
&=
G_2(Q_2^{2}) \mathcal{D}^{\lambda\lambda'}(\cos\hat{\theta}_1,\hat{\phi}_1), \\
G_2(Q_2^{2})
&=\frac{Q_2^2}{16 \pi^2}\sqrt{ 1-\frac{4m_{\pi}^2}{Q_2^2} }, 
\end{align}
with
\begin{align}
\mathcal{D}^{\lambda\lambda'}(\cos\theta,\phi)
=\left(\begin{array}{ccc}
\frac{\sin^2\theta}{2}				&\frac{-\sin^2\theta e^{2i\phi}}{2}	&\frac{-\sin2\theta e^{i\phi}}{2\sqrt{2}}\\[8pt]
\frac{-\sin^2\theta e^{-2i\phi}}{2}	&\frac{\sin^2\theta}{2}			&\frac{\sin2\theta e^{-i\phi}}{2\sqrt{2}}\\[8pt]
\frac{-\sin2\theta e^{-i\phi}}{2\sqrt{2}}	&\frac{\sin2\theta e^{i\phi}}{2\sqrt{2}} &\cos^2\theta\\
\end{array}\right). \label{Dll}
\end{align}
Here $\lambda(\lambda') = +1,-1,0$ from the top rows (the left columns), 
and the diagonal elements are simply the squared of the $d$-functions, $d^{J=1}_{\lambda,0}(\theta)$. 

Let us define the density matrix of the vector resonance production after integrating out the momenta of the two  unmeasurable neutrinos as 
\begin{align}
\mathcal{P}^{\lambda\lambda'}(q^2, E_V, \cos\theta_V, Q_n^2) 
=
\left. 
\frac{d\Gamma_V^{\lambda\lambda'}}{dq^2 dE_V d\cos\theta_{V}} \right\vert_{m_V^2 = Q_n^2}. 
\end{align}
We can now write the differential decay rate for all measurable momenta as
\begin{align}
\frac{d\Gamma_2}{dq^2 dE_V d\cos\theta_{V}  dQ_2^2  d\cos\hat{\theta}_1 d\hat{\phi}_1}
&=
\sum_{\lambda,\lambda'}
\mathcal{P}_2^{\lambda \lambda '}(q^2, E_V, \cos\theta_V, Q_2^2)
\mathcal{D}^{\lambda \lambda '}(\cos\hat{\theta}_1 ,\hat{\phi}_1), 
\end{align}
where 
\begin{align}
\mathcal{P}_2^{\lambda \lambda '}
&=
\mathcal{P}^{\lambda \lambda '}(q^2, E_V, \cos\theta_V, Q_2^2)
\frac{1}{2\pi} |F_2(Q_2^2)|^2 G_2(Q_2^2). 
\end{align}
From Eq. (\ref{Dll}), we obtain the following equation: 
\begin{align}
\frac{d\Gamma_2}{dq^2 dE_V d\cos\theta_{V}  dQ_2^2  d\cos\hat{\theta}_1 d\hat{\phi}_1}
&=
(\mathcal{P}_2^{++}+\mathcal{P}_2^{--})\frac{\sin ^2 \hat{\theta}_1}{2}+\mathcal{P}_2^{00} \cos ^2 \hat{\theta}_1 \label{dist2}\\
&-{\rm Re}(\mathcal{P}_2^{+-}) \sin^2\hat{\theta}_1 \cos 2\hat{\phi}_1
-{\rm Re}(\mathcal{P}_2^{+0}-\mathcal{P}_2^{-0})
	\frac{\sin2\hat{\theta}_1 \cos \hat{\phi}_1}{\sqrt{2}}\nonumber \\
&+{\rm Im}(\mathcal{P}_2^{+-}) \sin^2\hat{\theta}_1 \sin 2\hat{\phi}_1
  +{\rm Im}(\mathcal{P}_2^{+0}+\mathcal{P}_2^{-0})
	\frac{\sin 2\hat{\theta}_1 \sin \hat{\phi}_1}{\sqrt{2}}. \nonumber 
\end{align}
The differential decay rate of the CP conjugate process
$B \to \overline{D} \pi^+ \pi^0 \nu_{\tau} \overline{\nu}_{\tau}$, where all the three
momenta are reversed, differ from Eq. (\ref{dist2}) only by the
last two terms:
\begin{align}
\frac{ d\Gamma_2 - d\overline{\Gamma}^{\rm P}_2 }
     { d q^2 d E_V d\cos\theta_V d Q_2^2 d\cos\hat{\theta}_1 d\hat{\phi}_1 }
&=
 2 {\rm Im}( \mathcal{P}_2^{+-} )\sin^2\hat{\theta}_1 \sin 2\hat{\phi}_1 \label{asym_2terms}\\
&+
 2 {\rm Im}( \mathcal{P}_2^{+0} + \mathcal{P}_2^{-0} ) \frac{\sin 2\hat{\theta}_1 \sin\hat{\phi}_1}{\sqrt{2}}. \nonumber
\end{align}
Here $\hat{\theta}_1$ and $\hat{\phi}_1$ for $\pi^+$ are defined exactly the same as
for $\pi^-$ in Eq. (\ref{p1def}).  It should be noted here that the last two
terms in Eq. (\ref{dist2}) change sign under parity transformation.  The CP
violating term hence appears as $P$-odd distribution in the process
$\overline{B} \to D \pi^- \pi^0 \nu_{\tau} \overline{\nu}_{\tau}$, 
and it reverses the sign for the CP-conjugate process
$B \to \overline{D} \pi^+ \pi^0 \nu_{\tau} \overline{\nu}_{\tau}$. 

Actually, the sensitivity of ${\rm Im}( \mathcal{P}_2^{+-} )$ to the imaginary parts 
of the Wilson coefficients ($C_{V,S,T}$ of Eq. (\ref{GammaNP}), or more presicely
speaking, the phase of $\widetilde{C}_{T,S}$ in Eq. (\ref{CXpara}), since the phase
of $C_V$ is unobservable in our approximation) turns out to be very
small as compared with that of ${\rm Im}( \mathcal{P}_2^{+0} + \mathcal{P}_2^{-0} )$.  Therefore,
we discuss only the asymmetry due to the last term of Eq. (\ref{asym_2terms}).
We define the $q^2$ and $E_V$ dependent asymmetry distribution after
integrating over all the other kinematical variables as follows:
\begin{align}
A_2(q^2,E_V)
&\equiv
\frac{1}{\Gamma_2 + \overline{\Gamma}}_2
	\int d\cos\theta_{V}  dQ_2^2
	\left(  \int_0^1 - \int_{-1}^0  \right)d\cos\hat{\theta}_1
	\left(  \int_0^{\pi} - \int_{\pi}^{2\pi}  \right)d\hat{\phi}_1 \label{A2}\\
&~~~~~~~~~\times \frac{d\Gamma_2 - d\overline{\Gamma}_2^{\rm P}}
	{dq^2 dE_V d\cos\theta_{V}  dQ_2^2  d\cos\hat{\theta}_1 d\hat{\phi}_1}, \nonumber\\
&=\frac{32}{3\sqrt{2}(\Gamma_2 + \overline{\Gamma}_2)}
	\int d\cos\theta_{V}  dQ_2^2 ~ 
	{\rm Im}(\mathcal{P}_2^{+0}+\mathcal{P}_2^{-0}), 
\end{align}
where $\Gamma_2$ and $\overline{\Gamma}_2$ are the total decay rate of the process
$\overline{B} \to D \pi^- \pi^0 \nu_{\tau} \overline{\nu}_{\tau}$ and 
$B \to \overline{D} \pi^+ \pi^- \nu_{\tau} \overline{\nu}_{\tau}$.

%
%
%
\subsection{CP violating observable in three-pion decay}
\label{CPn3}
The amplitude of $a_1 \to 3 \pi$ is
\begin{align}
\mathcal{M}_3^{\lambda}
&=
\frac{4}{3f_{\pi}} \epsilon^{\alpha}(Q_3,\lambda)[
  F_2(P_{12}^2)(p_1-p_2)_{\alpha}
+F_2(P_{13}^2)(p_1-p_3)_{\alpha}
].
\end{align}
The contribution of this amplitude becomes large when $P_{12}^2$ or $P_{13}^2$ are near
the pole of the $\rho$ meson. 
When both $P_{12}$ and $P_{13}$ are near the pole, 
the amplitude is enhanced. 
Roughly the half of the distribution fall in the region. 
On the double $\rho$ meson pole, the $a_1 \to 3\pi$ decay amplitude is written as  
\begin{align}
\mathcal{M}_3^{\lambda} |_{P_{12}^2=P_{13}^2=m_{\rho}^2}
&=
\frac{4}{3f_{\pi}} \epsilon^{\alpha}(Q_3,\lambda)[
    F_2(m_{\rho}^2)(p_1-p_2)_{\alpha}
  +F_2(m_{\rho}^2)(p_1-p_3)_{\alpha}], \\
&\sim
\frac{4}{3f_{\pi}} \epsilon^{\alpha}(Q_3,\lambda)[
    \frac{i m_{\rho}}{\Gamma_{\rho}}(p_1-p_2)_{\alpha}
  +\frac{i m_{\rho}}{\Gamma_{\rho}}(p_1-p_3)_{\alpha}], \\
  &=
\frac{4}{3f_{\pi}} \frac{2i m_{\rho}}{\Gamma_{\rho}} \epsilon^{\alpha}(Q_3,\lambda)
    (p_1-\frac{p_2+p_3}{2})_{\alpha} \label{M3mom}. 
\end{align}
This has the same momentum structure as that of $V \to 2\pi$. 
Furthermore, under the CP transformation, both $\vec{p}_1$ and 
$(\vec{p}_2+\vec{p}_3)/2$ reverse the sign, because we assign 
$p_2$ and $p_3$ for two identical $\pi$'s consistently for all 3$\pi$ 
decay modes; see Eqs. (\ref{katei}d-e) for $a_1^-$ and Eqs. (\ref{CPkatei}d-e) for $a_1^+$. 
Therefore we can define the CP asymmetry exactly the same way 
as in the $V \to 2\pi$ case, by retaining only the sum of the 
two identical $\pi$'s momenta in the 3-body phase space. 
This definition is applicable even when the one of $P_{12}^2$ and $P_{13}^2$ is near
the pole of the $\rho$ meson as shown in appendix \ref{single_pole}. 

\begin{figure}[t]
\begin{center}
\includegraphics[width=9.0cm, bb=0 0 435 292]{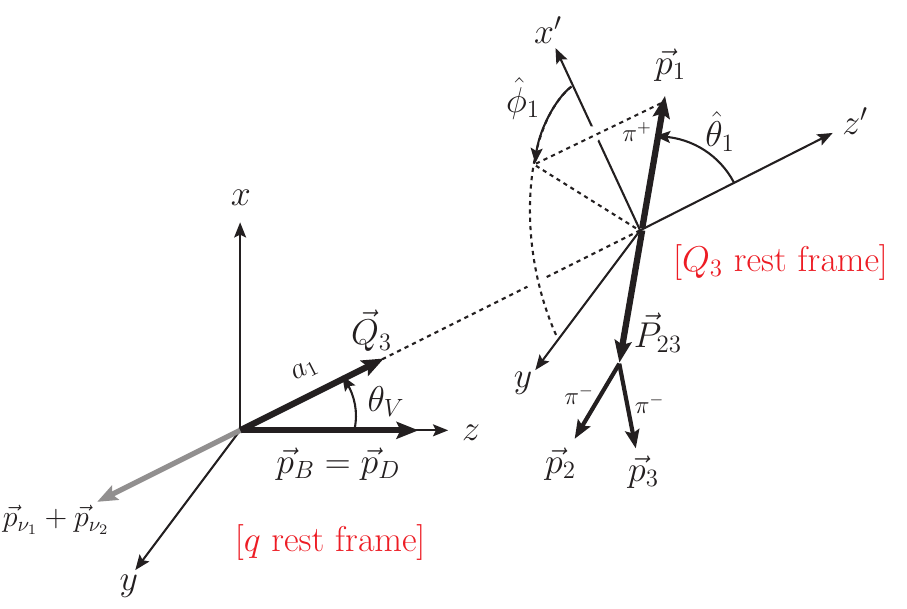}
\caption{{\footnotesize The kinematics in the case of $V\to 3\pi$.}}
\label{n3kin}
\end{center}
\end{figure}

We parametrize the momenta $Q_3$ in the $q$ rest frame as
\begin{align}
Q_3&=(E_V,|\vec{p}_V|\sin\theta_V,0,|\vec{p}_V|\cos\theta_V), 
\end{align}
where $E_V$ is the sum of three pions energies. The parametrization of $p_B$ and $p_D$, and definition of axis directions are the same as those of $V \to 2\pi$, see Eqs. (\ref{pBdef2pi}-\ref{Q2def2pi}). 
The opening angle made by $\vec{p}_D$ and $\vec{Q}_3$ is denoted by $\theta_V$. 
Then, we parametrize the momenta $p_1$ and $P_{23}$ in the $Q_3$ rest frame ($ \vec{p}_1+\vec{p}_2+\vec{p}_3=\vec{0}$) as
\begin{align}
p_1&=\frac{\sqrt{Q_3^2}}{2}
   (1+\frac{m_{\pi}^2}{Q_3^2}-\frac{P_{23}^2}{Q_3^2},+\beta_1' \sin\hat{\theta}_1\cos\hat{\phi}_1,
   	+\beta_1' \sin\hat{\theta}_1\sin\hat{\phi}_1,+\beta_1' \cos\hat{\theta}_1)\\
P_{23}&=\frac{\sqrt{Q_3^2}}{2}
   (1+\frac{P_{23}^2}{Q_3^2}-\frac{m_{\pi}^2}{Q_3^2},-\beta_1' \sin\hat{\theta}_1\cos\hat{\phi}_1,
   	-\beta_1' \sin\hat{\theta}_1\sin\hat{\phi}_1,-\beta_1' \cos\hat{\theta}_1)
\end{align}
where $\beta_1'=\bar{\beta}\left( \frac{m_{\pi}^2}{Q_3^2} , 
\frac{P_{23}^2}{Q_3^2} \right)$ and $P_{23} = p_2 + p_3$. 
In this frame, we define $z'$-axis as the direction of $\vec{Q}_3$ in the $q$ rest frame, and $y'$-axis as the direction of the $y$-axis. 
The polar and azimuthal angle of $\vec{p}_1$, $\hat{\theta}_1$ and $\hat{\phi}_1$, 
are parameterized as shown in Fig. \ref{n3kin}. 
With this approximation, the decay density matrix 
has the same form as in Eq. (\ref{Dll}) for a given $P_{23}^2$ after 
integrating over the internal phase space that keeps $P_{23}$. 
We therefore arrive at the CP asymmetry for the 3$\pi$ process 
which has the same form as in Eq. (\ref{asym_2terms}):
\begin{align}
A_3(q^2,E_V)
&\equiv
\frac{1}{\Gamma_3 + \overline{\Gamma}_3}
	\int d\cos\theta_{V}  dQ_3^2
	\left(  \int_0^1 - \int_{-1}^0  \right)d\cos\hat{\theta}_{1}
	\left(  \int_0^{\pi} - \int_{\pi}^{2\pi}  \right)d\hat{\phi}_1 \label{A3}\\
&~~~~~~~~~\times \frac{d\Gamma_3 - d\overline{\Gamma}_3^{\rm P}}
	{dq^2 dE_V d\cos\theta_{V}  dQ_3^2  d\cos\hat{\theta}_{1} d\hat{\phi}_{1}} \nonumber. 
\end{align}
where $\Gamma_3$ and $\overline{\Gamma}_3$ denote the total decay width of 
$\overline{B} \to D (3\pi) \nu_{\tau} \overline{\nu}_{\tau}$ and 
$B \to \overline{D} (3\pi) \nu_{\tau} \overline{\nu}_{\tau}$. 
In the above equation, integration over the invariant mass $P_{23}^2$ and the internal 
two body $(p_2+p_3)$ phase space has been suppressed for brevity. 
Although we can define more sophisticated CP asymmetry for the 
3$\pi$ mode by making full use of the 3-body decay kinematics, 
see e.g. Ref. \cite{Choi:1994ch}, we find that the above simple 
asymmetry has sufficiently strong sensitivity to CP violation, 
as shown in the next section. 
\section{Numerical results}
\label{Numerical}
\subsection{Model independent analysis}
\label{Model_independent_analysis}
\begin{figure}[t]
\begin{center}
\includegraphics[width=8.0cm, bb=0 0 237 184]{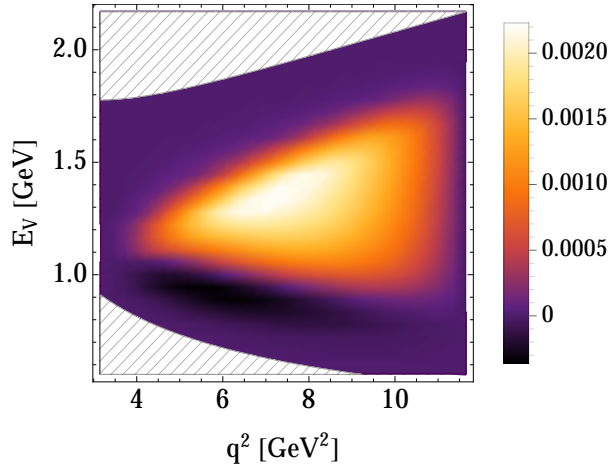}
\caption{{\footnotesize 
Density plot of the CP asymmetry function $A_2(q^2,E_V)$ in $C_S=i$ and $C_V=C_T=0$. 
The shaded areas are prohibited kinematically. }}
\label{A21}
\end{center}
\end{figure}
We estimate the sensitivity of the distributions $A_2$ and $A_3$ to the imaginary parts of new physics Wilson coefficients. 

In Fig. \ref{A21}, we show the density plot of the CP asymmetry function $A_2(q^2,E_V)$ in $C_S=i$ and $C_V=C_T=0$. 
The shaded areas are prohibited kinematically. 
The shape of density distributions are same for other value of $C_S$. 
In the tensor case, the distribution slightly differs from that of $C_S$. 
In the case of ${\rm Im}(C_{S,T})=0$, the distribution is zero on all phase space. 
\begin{figure}[t]
\begin{center}
\includegraphics[width=7.0cm, bb=0 0 280 179]{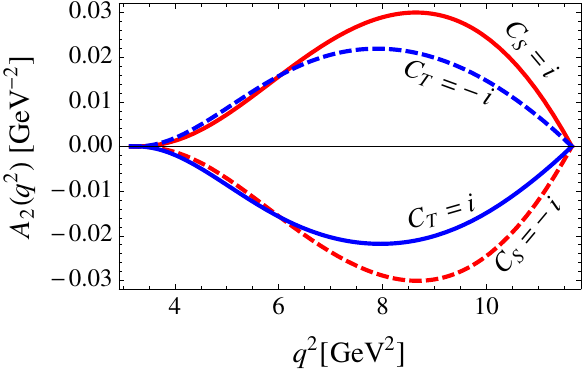}~~~~
\includegraphics[width=7.0cm, bb=0 0 280 179]{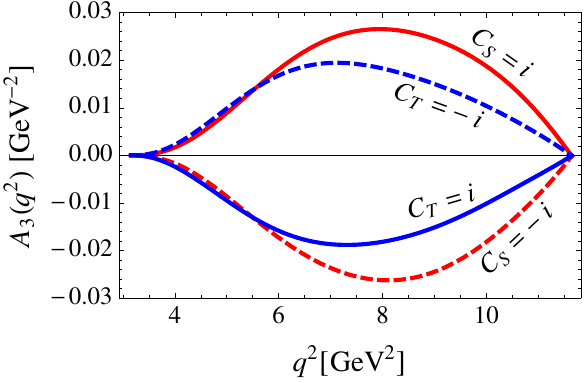}
\caption{{\footnotesize 
The CP asymmetry distributions $A_n(q^2)$ = $\int dE_V A_n(q^2,E_V)$
for the $2\pi~(n=2)$ mode (left), and for the $3\pi~(n=3)$ mode (right).
The red-solid curves show results for $(C_V,C_S,C_T)=(0,i,0)$,
the red-dashed curves are for $(0,-i,0)$,
the blue-solid curves are for $(0,0,i)$, and
the blue-dashed curves are for $(0,0,-i)$.
}}
\label{An2}
\end{center}
\end{figure}

In Fig. \ref{An2}, we show the distributions $A_n(q^2)$ = $\int dE_V A_n(q^2,E_V)$
for the $2\pi~(n=2)$ mode (left), and for the $3\pi~(n=3)$ mode (right).
The red-solid curves show results for $(C_V,C_S,C_T)=(0,i,0)$,
the red-dashed curves are for $(0,-i,0)$,
the blue-solid curves are for $(0,0,i)$, and
the blue-dashed curves are for $(0,0,-i)$.
Shapes of the distributions for the scalar effect are different from those for the tensor effect.  
Therefore, we can discriminate the type of the new physics interaction inducing CP violation 
by using these $q^2$ distributions. 
When the sign of the imaginary part change, the sign of the distributions also change. 
Note that the sign of distributions for the 2$\pi$ and 3$\pi$ mode are same for each models. 
So, unwished cancellation doesn't happen 
even if 2$\pi^0$'s in the $a_1 \to \pi^{\pm}\pi^0\pi^0 $ decay are not resolved, 
then it should contribute to the $\rho \to \pi^{\pm}\pi^0$ mode analysis. 
%
\begin{figure}[t]
\begin{center}
\includegraphics[width=6.0cm, bb=0 0 250 256]{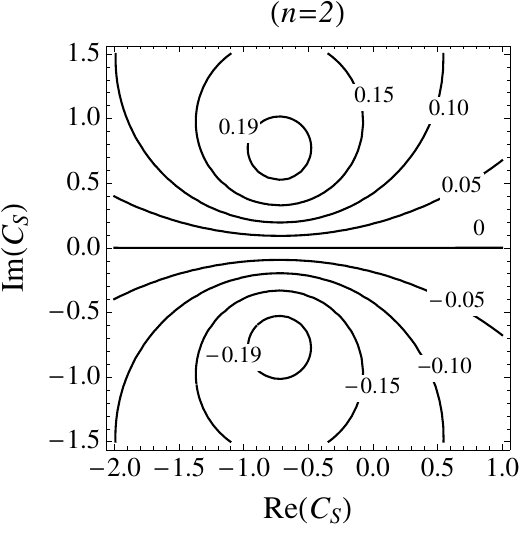}~~~
\includegraphics[width=6.0cm, bb=0 0 250 256]{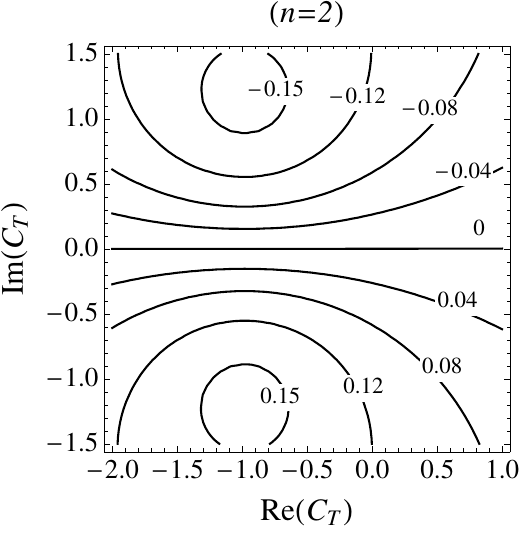}
\caption{{\footnotesize 
Contour plots of the integrated CP asymmetry, 
$A_2 = \int d q^2 d E_V A_2(q^2,E_V)$, on the plane of complex 
coefficients $C_S$ (left) and $C_T$ (right), when all the other 
Wilson coefficients are set to zero. 
}}
\label{A23}
\end{center}
\end{figure}
\begin{figure}[t]
\begin{center}
\includegraphics[width=6.0cm, bb=0 0 250 256]{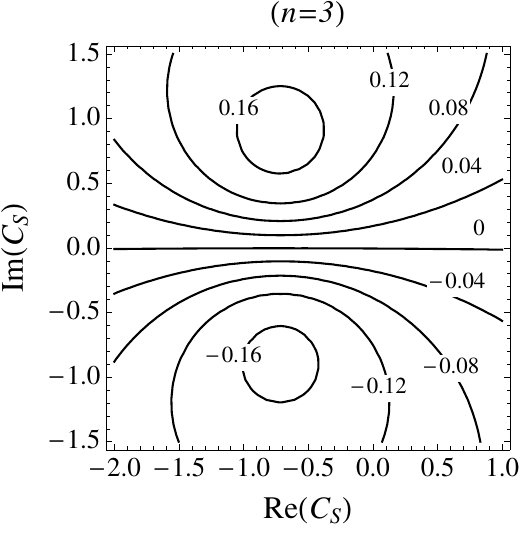}~~~
\includegraphics[width=6.0cm, bb=0 0 250 256]{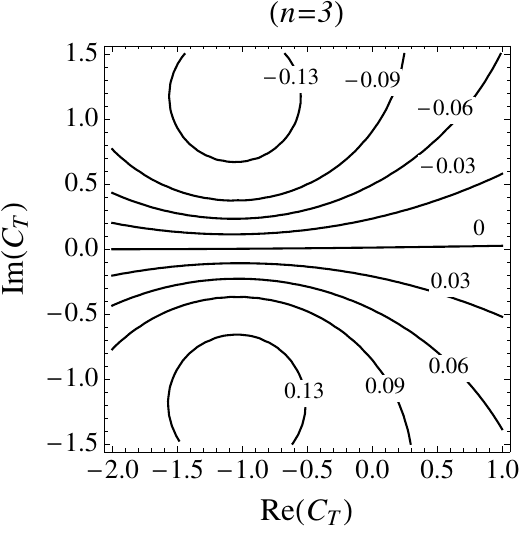}
\caption{{\footnotesize 
Contour plots of the integrated CP asymmetry, 
$A_3 = \int d q^2 d E_V A_3(q^2,E_V)$, on the plane of complex 
coefficients $C_S$ (left) and $C_T$ (right), when all the other Wilson coefficients are set to zero. 
}}
\label{A33}
\end{center}
\end{figure}

In Fig. \ref{A23} and \ref{A33}, we show 
contour plots of the integrated CP asymmetry, 
$A_n = \int d q^2 d E_V A_n(q^2,E_V)$, on the plane of the complex 
coefficients $C_S$ (left) and $C_T$ (right), when all the other Wilson coefficients are set to zero. 
The typical values are about 0.1 on ${\rm Im}(C_{S,T})\simeq 0.5$ for both $A_2$ and $A_3$. 
There are other integrated observables, for example, $\tau$ and $D^*$ polarizations \cite{Tanaka:2012nw,Tanaka:1994ay,Tanaka:2010se} and forward-backward asymmetries \cite{Sakaki:2012ft,Sakaki:2013bfa}. 
The behavior of $A_n$ for the wilson coefficients are different from such observables. 
Therefore, $A_n$ are useful to make the constraints on the Wilson coefficients be tight. 
It's notable that $A_n$ decide the sign of the imaginary part of Wilson coefficients.

\subsection{Model analysis}
It has been shown that the leptoquark models and the 2HDM of type-III have a possibility to explain the present experimental datas. 
So, we examine predictions for the size of the CP violation for these models. 

\subsubsection{Leptoquark models}
Some leptoquark models have the parameter spase which explain the experimental datas \cite{Sakaki:2013bfa}. 
In this paper, we consider three leptoquark models as follows:
\begin{align}
\mathcal{L}_{R_2}
&=(h_{2L}^{ij} \overline{u}_{iR}L_{jL}+h_{2R}^{ij} \overline{Q}_{iL} i \sigma_2 l_{jR}) R_2, \\
\mathcal{L}_{S_1}
&=(g_{1L}^{ij} \overline{Q}_{iL}^c i \sigma_2 L_{jL}+g_{1R}^{ij} \overline{u}_{iR}^c l_{jR}) S_1, \\
\mathcal{L}_{V_2}
&=(g_{2L}^{ij} \overline{d}_{iR}^c \gamma^{\mu} L_{jL}+g_{2R}^{ij} \overline{Q}_{iL}^c \gamma^{\mu} l_{jR}) V_2
\end{align}
The quantum numbers of the leptoquarks are summarized in Table \ref{tab:LQ_numbers}. 
\begin{table}[t]
   \begin{center}
   \begin{tabular}{|c|c|c|c|}
      \hline
					& $R_2$	& $S_1$	& $V_2$	\\
      \hline
      spin				& 0		& 0		& 1		\\
      \hline
      $F=3B+L$			& 0		& $-2$	& $-2$	\\
      \hline
      $SU(3)_c$			& 3		& $3^*$	& $3^*$	\\
      \hline
      $SU(2)_L$			& 2		& 1		& 2		\\
      \hline
      $U(1)_{Y=Q-T_3}$	& 7/6	& 1/3	& 5/6	\\
      \hline
   \end{tabular}
   \caption{\footnotesize Quantum numbers of the $R_2$, $S_1$, and $V_2$ leptoquarks with $SU(3)_c\times SU(2)_L\times U(1)_Y$ invariant couplings.}
   \label{tab:LQ_numbers}
\end{center}
\end{table}
Then, the Wilson coefficients of these leptoquarks are given as
\begin{align}
C_{V_1}&=
	\frac{1}{2\sqrt{2}G_F V_{cb}}
	\sum_{k=1}^{3}
	V_{k3}
		\frac{g_{1L}^{k3}g_{1L}^{23*}}{2M_{S_1^{1/3}}^2},
		\\
C_{S_1}&=
	\frac{-1}{2\sqrt{2}G_F V_{cb}}
	\sum_{k=1}^{3}
	V_{k3}
		\frac{2g_{2L}^{k3}g_{2R}^{23*}}{M_{V_2^{1/3}}^2},
		\\
C_{S_2}&=
	\frac{-1}{2\sqrt{2}G_F V_{cb}}
	\sum_{k=1}^{3}
	V_{k3}
	\left(
		  \frac{h_{2L}^{23}h_{2R}^{k3*}}{2M_{R_2^{2/3}}^2}
		+\frac{g_{1L}^{k3}g_{1R}^{23*}}{2M_{S_1^{1/3}}^2}
		\right),\\
C_{T}&=
	\frac{-1}{2\sqrt{2}G_F V_{cb}}
	\sum_{k=1}^{3}
	V_{k3}
	\left(
		\frac{h_{2L}^{23}h_{2R}^{k3*}}{8M_{R_2^{2/3}}^2}
		-\frac{g_{1L}^{k3}g_{1R}^{23*}}{8M_{S_1^{1/3}}^2}
		\right).
\end{align}
It's interesting that the $R_2$ and $S_1$ leptoquark models produce the combinations of the scalar and tensor interactions. 
The favored imaginary parts of the product of couplings for these leptoquark models with the leptoquark mass of 1 TeV have been estimated in Ref. \cite{Sakaki:2013bfa} as
\begin{align}
&1.92 < |{\rm Im}(h_{2L}^{23}h_{2R}^{33*})| < 2.42,\hspace{10pt}(R_2~{\rm leptoquark})\label{best1}\\
&1.92 < |{\rm Im}(g_{1L}^{33}g_{1R}^{23*})| < 2.42,\hspace{12pt}(S_1~{\rm leptoquark})\label{best2}\\
&0.34 < |{\rm Im}(g_{2L}^{33}g_{2R}^{23*})| < 0.68.\hspace{14pt}(V_2~{\rm leptoquark})\label{best3}
\end{align}
\begin{figure}[t]
\begin{center}
\includegraphics[width=7.0cm, bb=0 0 280 177]{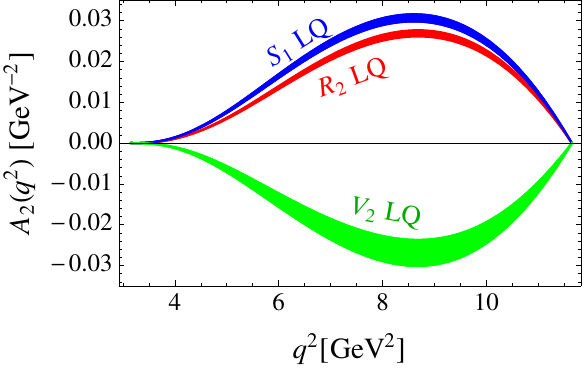}~~~~~%
\includegraphics[width=7.0cm, bb=0 0 280 179]{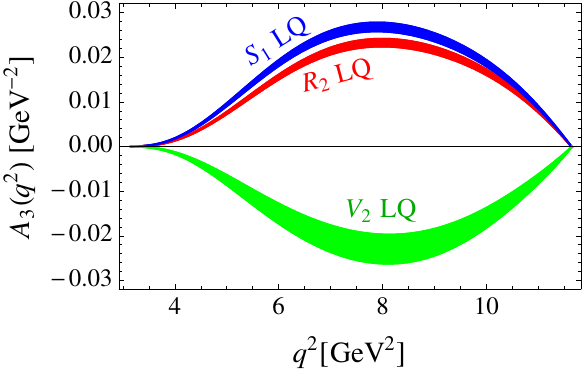}%
\caption{{\footnotesize 
The CP asymmetry distributions $A_n(q^2)$ = $\int dE_V A_n(q^2,E_V)$ 
for the product of couplings in the range given in Eqs. (\ref{best1}-\ref{best3}). 
The red (blue) [green] curve shows the distribution in the $R_2$ ($S_1$) [$V_2$] leptoquark model. 
We set the sign of the imaginary part of the coupling product for the $R_2$ ($S_1$) [$V_2$] leptoquark to $-$ $(-)$ $[+]$. 
}}
\label{LQAn2}
\end{center}
\end{figure}
In Fig. \ref{LQAn2}, we show the CP asymmetry distributions $A_n(q^2)$ = $\int dE_V A_n(q^2,E_V)$ 
for these couplings in Eqs. (\ref{best1}-\ref{best3}). 
The red (blue) [green] curve shows the distribution in the $R_2$ ($S_1$) [$V_2$] leptoquark model. 
We set the sign of the imaginary part of the coupling product for the $R_2$ ($S_1$) [$V_2$] leptoquark to $-$ $(-)$ $[+]$. 
Central values of  the integrated CP asymmetry, $A_n = \int d q^2 d E_V A_n(q^2,E_V)$, 
for these couplings are shown in Table \ref{tab:LQAn3}. 
The signs of $A_n$ are opposite ones of the imaginary part of the coupling products for each case. 

In fact, the allowed regions for the present experimental datas in $R_2$ leptoquark model exist only around the region in Eq. (\ref{best1}). 
So, if the CP violation is not observed, the $R_2$ leptoquark model would be completely excluded. 
\begin{table}[t]
   \begin{center}
   \begin{tabular}{|c|c|c|c|}
      \hline
					& $R_2$		& $S_1$		& $V_2$		\\
      \hline
      $A_2$		& $\mp 0.13$	& $\mp 0.15$	& $\mp 0.13$	\\
      \hline
      $A_3$		& $\mp 0.11$	& $\mp 0.13$	& $\mp 0.11$	\\
      \hline
   \end{tabular}
   \caption{\footnotesize 
Central values of  the integrated CP asymmetry, $A_n = \int d q^2 d E_V A_n(q^2,E_V)$, 
for the couplings in Eq. (\ref{best1}-\ref{best3}). 
The signs of $A_n$ are opposite ones of the imaginary part of the coupling products for each case. 
}
   \label{tab:LQAn3}
\end{center}
\end{table}

\subsubsection{2HDM of type-III}
\begin{figure}[t]
\begin{center}
\includegraphics[width=5.5cm, bb=0 0 220 210]{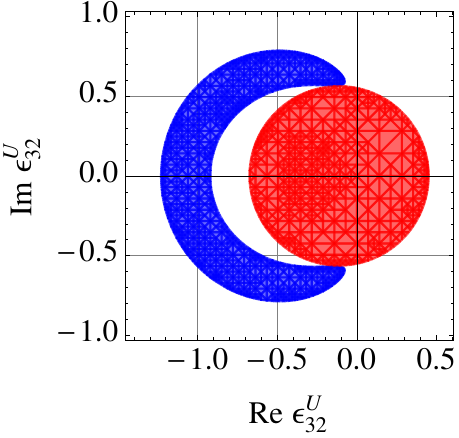}%
\caption{{\footnotesize 
The blue region shows the allowed region for the experimental datas of $R(D)$ and $R(D^*)$ on the parameter $\epsilon_{32}^{U}$ at 2$\sigma$ level. 
The red region shows the allowed region for the normalized $q^2$ distribution in $\overline{B}\to D\tau(\to \ell\nu\bar{\nu})\bar{\nu}$ with kinematic cut $m_{\rm miss}^2>1.5{\rm GeV}^2$ at 2$\sigma$ level. 
We set $\tan\beta=50$ and $m_{H^{\pm}}=500{\rm GeV}$. 
}}
\label{type3}
\end{center}
\end{figure}
\begin{figure}[t]
\begin{center}
\includegraphics[width=5.0cm, bb=0 0 250 175]{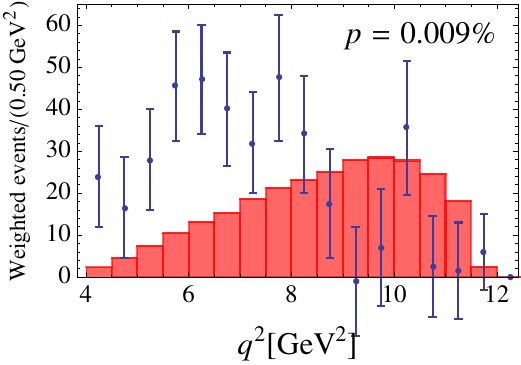}~~~%
\includegraphics[width=5.0cm, bb=0 0 250 175]{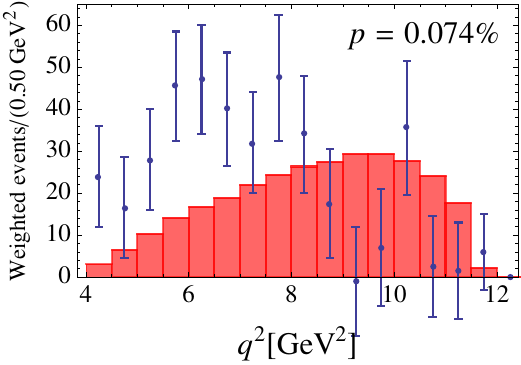}~~~%
\includegraphics[width=5.0cm, bb=0 0 250 175]{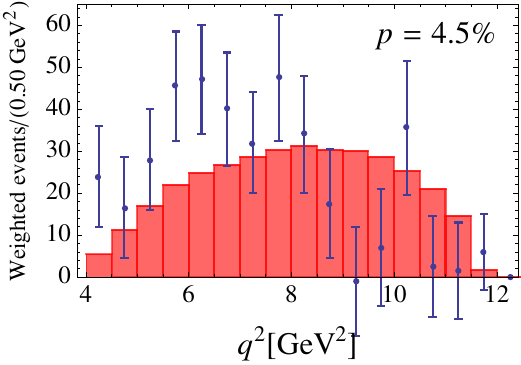}%
\caption{{\footnotesize 
The left (center) [right] histogram shows the theoretical prediction of the normalized $q^2$ distribution with cut for the parameter 
$\epsilon_{32}^U=-1$ ($\epsilon_{32}^U=-0.7\pm0.6i$) [$\epsilon_{32}^U=-0.13\pm0.57i$] 
and the experimental datas. 
The uncertainty on the data points includes the statistical uncertainties of data and simulation. 
The p-value for each parameters are shown in the figures. 
}}
\label{q2dist}
\end{center}
\end{figure}
The 2HDMs predict the charged Higgs bosons $H^{\pm}$ which affect the decay $B\to D\tau\nu$ at tree level. 
The Lagrangian including the charged Higgs boson in the 2HDM of type-III is written as:
\begin{align}
\mathcal{L}_{H^{\pm}}
&=
\frac{1}{v}
\left[
	\overline{u}_{iR}
	\left(
		\cot\beta~m_{u_i} \delta_{ij}
		-\frac{v}{\sin\beta}\epsilon_{ji}^{U*}
	\right)
	V_{jk}^{\rm CKM}d_{kL}\right. \\
&\hspace{17pt}
+\left. 
	\overline{u}_{iL}V_{ij}^{\rm CKM}
	\left(
		\tan\beta~m_{d_j} \delta_{jk}
		-\frac{v}{\cos\beta}\epsilon_{jk}^{D}
	\right)
	d_{kR}\right. \\
&\hspace{17pt}
+\left. 
	\overline{\nu}_{iL}
	\left(
		\tan\beta~m_{l_i} \delta_{ij}
		-\frac{v}{\cos\beta}\epsilon_{ij}^{E}
	\right)
	l_{jR}
\right]H^+
+{\rm H.c.}
\end{align}
where $v=174$GeV and $V^{\rm CKM}$ denote the Cabibbo-Kobayashi-Maskawa matrix. 
Then, the Wilson coefficients of this model are
\begin{align}
C_{S_1}
&=
-
\left(
	\tan\beta~m_b -\frac{v}{V_{cb}\cos\beta}\epsilon_{23}^D
\right)
\left(
	\tan\beta~m_{\tau} -\frac{v}{\cos\beta}\epsilon_{23}^{E*}
\right)
\frac{1}{m_{H^{\pm}}^2}, \\
C_{S_2}
&=
-
\left(
	\cot\beta~m_c -\frac{v}{V_{cb}\sin\beta}\epsilon_{32}^{U*}
\right)
\left(
	\tan\beta~m_{\tau} -\frac{v}{\cos\beta}\epsilon_{23}^{E*}
\right)
\frac{1}{m_{H^{\pm}}^2},
\end{align}
where $V_{cb}\equiv V^{\rm CKM}_{j3}$, and we use an approximation $V^{\rm CKM}_{j3}\simeq \delta_{j3}$. 
The off-diagonal elements of the matrices $\epsilon^U$, $\epsilon^D$, and $\epsilon^E$ induce the flavor-changing neutral current, so the constraints on these parameters are tight. 
Actually, the constraint on the parameters $\epsilon_{23}^{D}$ and $\epsilon_{23}^{E}$ are tight, so we neglect these parameters. 
However, the constraint on the $\epsilon_{32}^{U}$ are not so tight, and there is an allowed region on this parameter, which have been shown in Ref. \cite{Crivellin:2012ye}. 

In this paper, we set $\tan\beta=50$ and $m_{H^{\pm}}=500{\rm GeV}$. 
In Fig. \ref{type3}, the blue region shows the allowed region for the experimental datas of $R(D)$ and $R(D^*)$ on the parameter $\epsilon_{32}^{U}$ at 2$\sigma$ level. 
The red region shows the allowed region at 2$\sigma$ level for the normalized $q^2$ distribution in $\overline{B}\to D\tau(\to \ell\nu\bar{\nu})\bar{\nu}$ with kinematic cut $m_{\rm miss}^2>1.5{\rm GeV}^2$, which is measured by the BaBar collaboration \cite{Lees:2013uzd}. 
This kinematic cut is the same as the BaBar's one. 
In Fig. \ref{q2dist}, the left (center) [right] histogram shows the theoretical prediction of the normalized $q^2$ distribution with the cut for the parameter 
$\epsilon_{32}^U=-1$ ($\epsilon_{32}^U=-0.7\pm0.6i$) [$\epsilon_{32}^U=-0.13\pm0.57i$] 
and the experimental datas. 
The uncertainty on the data points includes the statistical uncertainties of data and simulation. 
The p-value for each parameters are shown in these figures. Please see Ref. \cite{q2dist_pre} for more detail about the $q^2$ distribution analysis. 

It doesn't seem that the 2HDM of type-III is consistent with those experimental results. 
However, the p-values for the normalized $q^2$ distribution on $\epsilon_{32}^U=-0.13\pm0.57i$ 
which are boundary of the red and blue regions 
are larger than the ones on the allowed region for $R(D^{(*)})$. 
So, we assume these parameters as benchmark points in the 2HDM type-III. 

\begin{figure}[t]
\begin{center}
\includegraphics[width=6.0cm, bb=0 0 250 170]{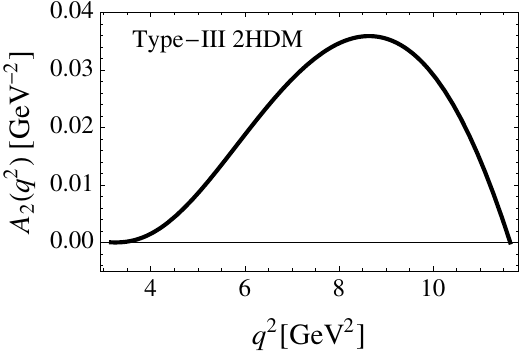}~~~~~~~
\includegraphics[width=6.0cm, bb=0 0 250 170]{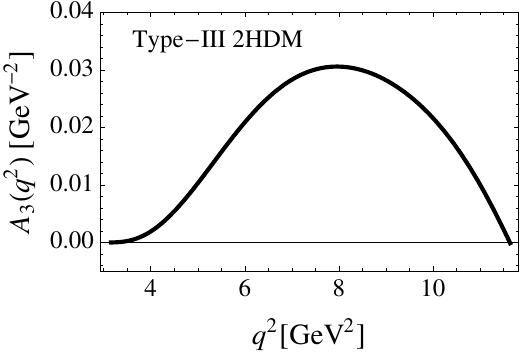}
\caption{{\footnotesize 
The left (right) figure shows the CP asymmetry distributions $A_2(q^2)$ ($A_3(q^2)$) on $\epsilon_{32}^U=-0.13-0.57i$. 
}}
\label{type3An2}
\end{center}
\end{figure}

In Fig. \ref{type3An2}, the left (right) figure shows the CP asymmetry distributions $A_2(q^2)$ ($A_3(q^2)$) on $\epsilon_{32}^U=-0.13-0.57i$, 
and the integrated CP asymmetry $A_n$ on $\epsilon_{32}^U=-0.13\pm0.57i$ are shown in table \ref{tab:type3An3}. 
The signs of $A_n$ are opposite ones of the imaginary part of the parameter $\epsilon_{32}^U$. 
\begin{table}[t]
   \begin{center}
   \begin{tabular}{|c|c|}
      \hline
					& type-III 2HDM\\
      \hline
      $A_2$		& $\mp 0.17$	\\
      \hline
      $A_3$		& $\mp 0.15$	\\
      \hline
   \end{tabular}
   \caption{\footnotesize 
The integrated CP asymmetry $A_n$ on $\epsilon_{32}^U=-0.13\pm0.57i$ are show in table \ref{tab:type3An3}. 
The signs of $A_n$ are opposite ones of the imaginary part of the parameter $\epsilon_{32}^U$. 
   }
   \label{tab:type3An3}
\end{center}
\end{table}

\section{Conclusion}
\label{Conclusion}
In this paper, we have constructed the CP asymmetry distribution $A_2(q^2, E_V)$  
for the decay processes $B\to D\tau(\to 2\pi)\nu_{\tau}$ and $A_3(q^2, E_V)$ for $B\to D\tau(\to 3\pi)\nu_{\tau}$
by using the polarization of the vector resonances produced by the tau lepton. 
Assuming the tau neutrinos to be left-handed, we have introduced the general effective Hamiltonian 
that contains all possible four-fermion operators of the lowest dimension for the $b\to c \tau \nu_{\tau}$. 
The two independent imaginary parts of the parameters $\widetilde{C}_S$ and $\widetilde{C}_T$ 
given by the Wilson coefficients of those operators induce the CP violation and nonzero 
distributions of $A_n(q^2, E_V)$. 
These two parameters are related with the non-standard-model interactions of scalar and tensor. 

We have examined the sensitivities of the $A_n(q^2, E_V)$ to the scalar and tensor interactions. 
Shapes of the CP aymmetric $q^2$ distribution, $A_n(q^2)=\int A_n(q^2, E_V)$ for the scalar interaction
are different from those of the tensor interaction. Therefore, we can discriminate the type of the new physics interaction inducing the CP violation using the distributions. 
The integrated CP asymmetries, $A_n = \int d q^2 d E_V A_n(q^2,E_V)$, 
are also sensitive to ${\rm Im}(\widetilde{C}_{S,T})$, and
take about 0.1 typically on ${\rm Im}(C_{S,T})\simeq 0.5$ for both 2$\pi$ and 3$\pi$ decays. 

Some new physics models have been considered to explain the discrepancy 
between the present experimental datas and the SM predictions, 
and would induce the CP violation. 
We have examined expected the CP violation for three leptoquark models. 
The favored imaginary parts of the coupling products in the leptoquark models have been known, 
and we have estimated the $A_n(q^2)$ and $A_n$ for those couplings. 
The allowed regions for the present experimental datas in the $R_2$ leptoquark model only exist on the nonzero imaginary part. 
So, if the CP violation in $B\to D\tau \nu$ is not observed, the $R_2$ leptoquark model would be completely excluded. 

We have also discussed the 2HDM of type-III, 
and indicated the inconsistency of this model by using the experimental datas for ratios of the measured branching fractions  $R(D^{(*)})$ and the normalized $q^2$ distribution in $\overline{B}\to D\tau\bar{\nu}$. 
Assuming parameters $\epsilon_{32}^U=-0.13\pm0.57i$ as benchmark points in this model, 
we have estimated the $A_n(q^2)$ and $A_n$ in the type-III 2HDM. 

In conclusion, 
the past measurements for  $B\to D \tau \nu$ in the BaBar and Belle have used only the leptonic tau decay modes, $\tau\to \ell \nu \overline{\nu} $. 
An analysis using the hadronic tau decay modes in semi-tauonic B decays would be difficult, 
however must give a lot of informations for the tau lepton polarization. 
We hope for successful measurements of the hadronic channels in Belle-II and LHCb.

\section*{Acknowledgements}
We thanks Ryoutaro Watanabe for very helpful discussions, 
and also Toru Goto, Junya Nakamura, and Minoru Tanaka for useful comments. 
This work is supported by 
the Grant-in-Aid for Scientific Research from the Ministry of Education, Science, Sports, 
and Culture (MEXT), Japan. (25400287, and 23104006).

\appendix
\section{Single $\rho$-meson-pole approximation}
\label{single_pole}
In Sec. \ref{CPn3} , we have explained that the decay density matrix for $V \to 3\pi$ has the same form as for $V \to 2\pi$ in Eq. (\ref{Dll}) when $P_{12}^2$ and $P_{13}^2$ are both near the pole of $\rho$ meson. 
In this appendix, we show that the correspondence between $V \to 3\pi$ and $V \to 2\pi$ is also applicable approximately 
even when the only one of $P_{12}^2$ and $P_{13}^2$ is near the invariant mass squared of $\rho$ meson. 
The magnitude of the distribution on this phase space is comparable with 
the one on the double $\rho$ meson pole which is discussed in Sec. \ref{CPn3}. 

The first (second) term in the square bracket in Eq. (\ref{M3}) is dominant when $P_{12}^2$ ($P_{13}^2$) is near the pole and 
$P_{13}^2$ ($P_{12}^2$) is far away from the pole. 
The decay density matrix distribution for $V \to 3\pi$ on these phase space regions is written approximately as 
\begin{align}
d\Gamma_3^{\lambda \lambda'}
\simeq
	\int
	\frac{16 \pi  m_{\rho }}{9 f_{\pi }^2 \Gamma _{\rho }}
	[
	&d\Phi_3
		\delta(P_{12}^2 - m_{\rho}^2)
		\epsilon_{\alpha }   (Q_3,\lambda) (p_1-p_2)^{\alpha}
		\epsilon_{\alpha'}^*(Q_3,\lambda') (p_1-p_2)^{\alpha'}\\
      &+(2 \to 3)]. \nonumber
\end{align}
We assume that the momentum squared $P_{13}^2$ ($P_{12}^2$) in the first (second) term 
which is not near the pole is significantly smaller than $m_{\rho}^2$. 
After integrating over the momentum squared $P_{23}^2$ and 
two body $(p_2 + p_3)$ phase space for massless pion's, 
we obtain the following equation:
\begin{align}
\frac
{d\Gamma_3^{\lambda \lambda'}}
{d\cos\hat{\theta}_1  d\hat{\phi}_1}
&\simeq
	G_3(Q_3^2) \widetilde{\mathcal{D}}^{\lambda\lambda'}(\cos\hat{\theta}_1,\hat{\phi}_1), \\
G_3(Q_3^{2})
&=
	\frac{(1+r) m_{\rho}^3}{576 \pi^3 f_{\pi}^2 \Gamma_{\rho} }, 
\end{align}
with
\begin{align}
\widetilde{\mathcal{D}}^{\lambda\lambda'}(\cos\theta,\phi)
=
	\frac{1}{3\delta}
	\left(\begin{array}{ccc}
	\frac{\sin^2\theta}{2} + \delta		&\frac{-\sin^2\theta e^{2i\phi}}{2}	&\frac{-\sin2\theta e^{i\phi}}{2\sqrt{2}}\\[8pt]
	\frac{-\sin^2\theta e^{-2i\phi}}{2}	&\frac{\sin^2\theta}{2} + \delta		&\frac{\sin2\theta e^{-i\phi}}{2\sqrt{2}}\\[8pt]
	\frac{-\sin2\theta e^{-i\phi}}{2\sqrt{2}}	&\frac{\sin2\theta e^{i\phi}}{2\sqrt{2}} &\cos^2\theta + \delta	\\
	\end{array}\right), 
\end{align}
where $\delta = (1+r)/7r^2$ and $r = m_{\rho}^2 / Q_3^2$. 
The off-diagonal elements of the decay density matrix $\widetilde{\mathcal{D}}^{\lambda\lambda'}$ 
is proportional to those for $V \to 2\pi$ in Eq. (\ref{Dll}). 
The differential decay rate $d\Gamma_3$ has similar form as $d\Gamma_2$ in Eq. (\ref{dist2}) due to the above equations. 
Using the CP asymmetry defined in Eq. (\ref{A3}), therefore, 
we can measure the coefficient function ${\rm Im}(\mathcal{P}^{+0} + \mathcal{P}^{-0})$ 
that measures CP violation. 



\end{document}